\documentstyle[lprocl,11pt,epsfig,floatfig]{article}

\begin{document}
\pagestyle{empty}

\initfloatingfigs
\setcounter{equation}{0}
\setcounter{figure}{0}

\begin{flushleft}
\Large
{SAGA-HE-101-96  
   \hfill May 20, 1996}  \\
\end{flushleft}
 
\vspace{2.5cm}
 
\begin{center}
 
\huge{{\bf Nuclear modification}} \\
\vspace{0.3cm}

\huge{{\bf of the flavor asymmetry $\bf\overline u - \overline d$}} \\
 
\vspace{1.5cm}
 
\LARGE
{S. Kumano $^*$ }         \\
 
\vspace{0.8cm}
  
\LARGE
{Department of Physics}         \\
 
\vspace{0.1cm}
 
\LARGE
{Saga University}      \\
 
\vspace{0.1cm}

\LARGE
{Saga 840, Japan} \\

\vspace{2.0cm}
 
\Large
{Talk given at the XIV International Conference} \\

\vspace{0.3cm}

{on Particles and Nuclei} \\

\vspace{0.7cm}

{Williamsburg, U.S.A., May 22 -- 28, 1996 (talk on May 27, 1996)}  \\
 
\end{center}
 
\vspace{1.3cm}

\vfill
 
\noindent
{\rule{6.cm}{0.1mm}} \\
 
\vspace{-0.2cm}
\large
\noindent
{* Email: kumanos@cc.saga-u.ac.jp. 
   Information on his research is available}  \\

\vspace{-0.45cm}
\noindent
{at http://www.cc.saga-u.ac.jp/saga-u/riko/physics/quantum1/structure.html} \\

\vspace{-0.45cm}
\noindent
\large
{or at ftp://ftp.cc.saga-u.ac.jp/pub/paper/riko/quantum1.} \\

\vspace{1.0cm}

\vspace{-0.5cm}
\hfill
{to be published in proceedings}

\vfill\eject
\normalsize



\title{NUCLEAR MODIFICATION OF THE FLAVOR ASYMMETRY 
        $\bf \overline u-\overline d$}

\author{ S. KUMANO $^*$ }

\address{ Department of Physics, Saga University \\
          Honjo-1, Saga 840, Japan} 


\maketitle\abstracts{
In order to test the NMC finding of flavor asymmetry 
$\overline u - \overline d$ in the nucleon, existing Drell-Yan data 
for the tungsten target are often used. 
However, we have to be careful in comparing nuclear data
with the nucleon ones.
We investigate whether there exists significant nuclear modification
of the $\overline u - \overline d$ distribution in a parton-recombination 
model. It should be noted that a finite $\overline u - \overline d$ 
distribution is theoretically possible in nuclei even if the sea is 
symmetric in the nucleon. In neutron-excess nuclei such as the tungsten,
there exist more $d$-valence quarks than $u$-valence quarks, so that
more $\overline d$-quarks are lost than $\overline u$-quarks due
to parton recombinations.
Our results suggest that the nuclear modification 
in the tungsten is a 2--10 \% effect on the 
NMC $\overline u - \overline d$ distribution.
The nuclear modification of the flavor asymmetry should be an interesting
topic in connection with ongoing Drell-Yan experiments.}


The New Muon Collaboration (NMC) suggested 
that the Gottfried sum rule should be violated in 1991.
It indicates $\bar d$ excess over $\bar u$ in the nucleon.
Since then, there have been efforts to investigate mechanisms of
creating a flavor asymmetric distribution $\bar u-\bar d$ 
in the nucleon.\cite{sk97}
In order to test the NMC finding, Drell-Yan experiments are in progress 
at Fermilab. On the other hand, there exist Drell-Yan data for various nuclear
targets, so that some people use, for example, tungsten data in 
investigating the flavor asymmetry. 
However, we have to be careful in comparing the NMC result
with the tungsten data because of possible nuclear medium effects.
In order to find whether such a comparison makes sense, 
we estimate a nuclear modification effect.\cite{sk95} 
It in turn could be found experimentally
by analyzing accurate Drell-Yan data in the near future. 

We investigate the $\bar u-\bar d$ distribution in the tungsten
nucleus. If isospin symmetry could be applied to parton distributions
in the proton and the neutron, the distribution per nucleon becomes
$x[\bar u(x)-\bar d(x)]_A = 
- \varepsilon x [\bar u(x)-\bar d(x)]_{proton}$
without considering nuclear modification. 
It is just the summation of proton and neutron contributions.
The neutron-excess parameter $\varepsilon$ is defined by
$\varepsilon =(N-Z)/(N+Z)$, and 
it is 0.196 for the tungsten $_{74}^{184} W_{110}$.
According to the above equation, the flavor distribution
should be symmetric ($[\bar u-\bar d]_W=0$) if it is symmetric
in the nucleon.
However, it is not the case in a parton-recombination model.

It is well known that shadowing phenomena occur
in nuclear structure functions at small $x$. 
One of the ideas for explaining
the shadowing is the recombination model.\cite{recomb}
In an infinite momentum frame, the longitudinal
localization size of a parton with momentum $xp_{_N}$
exceeds the average longitudinal
nucleon separation in a Lorentz contracted nucleus
[(2 fm)$M_A/P_A$=(2 fm)$m_{_N}/p_{_N}$]
in the small $x$ region ($x<0.1$).
Therefore, partons from different nucleons could interact,
and the interaction is called parton recombination.
We apply studies of this model in the structure function $F_2$
to the asymmetry $\bar u-\bar d$ in nuclei.

The flavor asymmetry could be created in the recombination model
in the following way.
In a neutron-excess nucleus ($\varepsilon >0$) such
as the tungsten, more $\bar d$ quarks are lost than $\bar u$ quarks
in the parton recombination process 
$\bar q q \rightarrow G$ because of
the $d$ quark excess over $u$ in the nucleus.
The $\bar q q \rightarrow G$ type
recombination processes produce positive contributions
at small $x$.
In the $(\bar u-\bar d)_N \ne 0$ case, 
the $\bar q(x) G\rightarrow \bar q$ process
is the dominant one kinematically at small $x$.
Its contribution to $\bar u(x)-\bar d(x)$ becomes negative
due to the neutron excess.
In the medium $x$ region, the $\bar q G\rightarrow \bar q(x)$
process becomes kinematically favorable. 
Because it produces
$\bar q$ with momentum fraction $x$, its contribution becomes
opposite to the one at small $x$.
These results are shown in Fig. \ref{fig:ub-db}.

\begin{floatingfigure}{7.0cm}
   \begin{center}
      \mbox{\epsfig{file=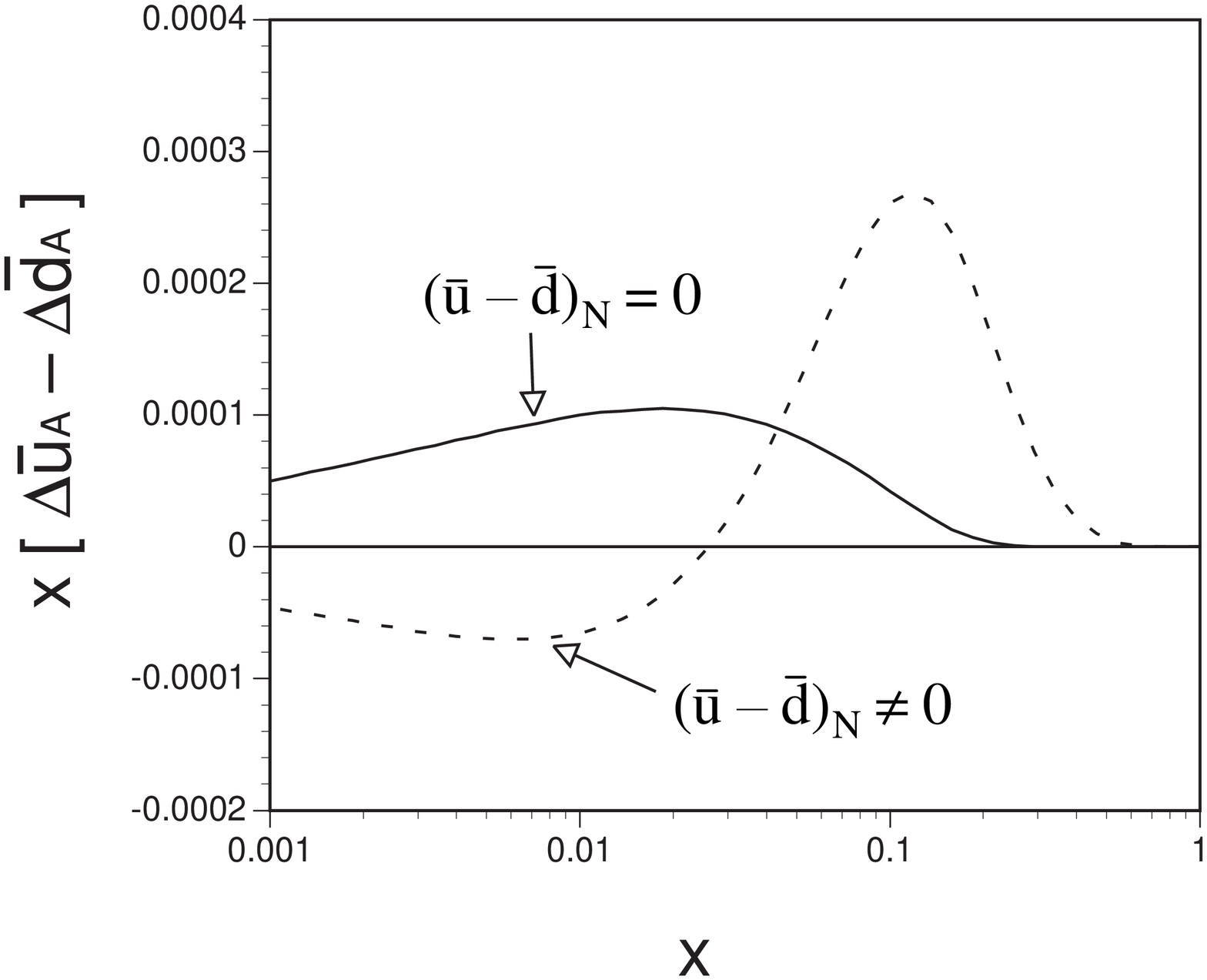,width=5.0cm}}
   \end{center}
 \vspace{-0.6cm}
\caption{\footnotesize Recombination contributions to $\bar u-\bar d$
                       in the tungsten nucleus.}
\label{fig:ub-db}
\end{floatingfigure}
\quad
We evaluate the recombination effects 
on the $\bar u-\bar d$ distribution at $Q^2$=4 GeV$^2$
in the tungsten nucleus $_{74}^{184} W_{110}$ 
($\varepsilon=0.196$).\cite{sk95}
Input parton distributions are those of the MRS-D0 (1993).
In the $(\bar u-\bar d)_N = 0$ case, 
$\Delta=0$ is taken in the MRS-D0 distributions.
Obtained results are shown in Fig. \ref{fig:ub-db}, 
where the solid curve shows recombination contributions to $\bar u-\bar d$
(per nucleon) in the tungsten nucleus,
and the dashed curve shows the result
in the $(\bar u-\bar d)_N \ne 0$ case.

We briefly comment on $Q^2$ dependence of our calculation.
Because the recombinations are higher-twist effects, there is
a $Q^2$ dependent factor $\alpha_s(Q^2)/Q^2$ in the contributions.
It may seem to be very large at small $Q^2$;
however, parton distributions $p(x,Q^2)$ are also modified.
As a consequence, the overall $Q^2$ dependence is not so significant. 
According to our estimate, there are merely factor-of-two differences
between the asymmetric distribution 
at $Q^2$=4 GeV$^2$ and the one at $Q^2 \approx 1$ GeV$^2$.
Considering this factor of two,
we find that the nuclear modification is of the order of 2\%--10\%
compared with the asymmetry suggested by the NMC.

Although our estimate is based on a special model, the results
indicate roughly several \% nuclear modification in the $\bar u-\bar d$
distribution of the tungsten nucleus.
Therefore, the existing tungsten Drell-Yan data 
should not be compared with the NMC flavor asymmetry
within several \% magnitude.
On the other hand, the modification itself is an interesting topic
for future theoretical and experimental studies.

\vspace{-0.3cm}
\section*{Acknowledgment}
\vspace{-0.2cm}
This research was partly supported by the Grant-in-Aid for
Scientific Research from the Japanese Ministry of Education,
Science, and Culture under the contract number 06640406.

\vspace{+0.3cm}

\noindent
{* Email: kumanos@cc.saga-u.ac.jp. 
   Information on his research is available}  \\

\vspace{-0.6cm}
\noindent
{\ \ \ 
at http://www.cc.saga-u.ac.jp/saga-u/riko/physics/quantum1/kumano.html.} \\


\end{document}